\documentclass[aps,llncs,preprint]{revtex4-2}
\usepackage{graphicx}
\usepackage{amsmath}
\usepackage{float}
\usepackage{color}

\begin{document}
	
\title{Spectroscopic evidence of mixed angular momentum symmetry in non-centrosymmetric Ru$_7$B$_3$}

\author{Soumya Datta$^1$, Aastha Vasdev$^1$, Ranjani Ramachandran$^1$, Soumyadip Halder$^1$, Kapil Motla$^2$, Anshu Kataria$^2$, Arushi$^2$, Rajeswari Roy Chowdhury$^2$, Ravi Prakash Singh$^2$, and Goutam Sheet$^1$}

\email{goutam@iisermohali.ac.in}

\affiliation{$^1$Department of Physical Sciences, Indian Institute of Science Education and Research Mohali, Sector 81, S. A. S. Nagar, Manauli, PO 140306, India}

\affiliation{$^2$Department of Physics, Indian Institute of Science Education and Research Bhopal, Bhopal 462066, India}

\begin{abstract}
	
{Superconducting crystals with lack of inversion symmetry can potentially host unconventional pairing. However, till date, no direct conclusive experimental evidence of such unconventional order parameters in non-centrosymmetric superconductors has been reported. In this paper, through direct measurement of the superconducting energy gap by scanning tunnelling spectroscopy, we report the existence of both $s$-wave (singlet) and $p$-wave (triplet) pairing symmetries in non-centrosymmetric Ru$_7$B$_3$. Our temperature and magnetic field dependent studies also indicate that the relative amplitudes of the singlet and triplet components of the order parameter change differently with temperature.}

\end{abstract}

\maketitle

In the BCS theory, it is assumed that the attractive interaction that leads to superconductivity is isotropic in momentum space\cite{Bardeen}. Consequently, the superconducting energy gaps of BCS superconductors show $s$-wave (orbital angular momentum, $l$=0) symmetry. In certain superconducting systems, the energy gap may become anisotropic in the momentum space, and show higher angular momentum symmetries like $p$-wave ($l$=1)\cite{Ott,Ueda,Maeno}, $d$-wave ($l$=2)\cite{Wollman,Tsuei,Tsuei2} etc. In certain other systems, existence of mixed angular momentum symmetry, where symmetries represented by different $l$ are mixed, have also been possible\cite{Steglich,Tokuyasu,Kouznetsov,Annett}. The physics of such non-$s$-wave superconductors are not understood within the BCS formalism. If the crystal structure of a superconductor lacks a centre of inversion symmetry, parity is not a good quantum number. In such a system an antisymetric spin-orbit coupling (ASOC) can exist. ASOC can in-principle remove the spin degeneracy of the Bloch states with same $k$ (crystal momentum), but opposite spins. In presence of ASOC, the orbital angular momentum and spin angular momentum do not remain good quantum numbers any longer. Here, Pauli's exclusion principle cannot restrict the symmetry of the Cooper pairs to be either purely even-parity singlet or odd-parity triplet. Therefore, a complex mixed angular momentum state becomes a possibility in a non-centrosymmetric superconductor\cite{Frigeri}. The unconventionality associated with such complex angular momentum symmetry of the superconducting order parameters might make the non-centrosymmetric superconductors (NCS) exhibit unusual behaviour in their electro-magnetic properties compared to the purely $s$-wave superconductors. For example, they may display unusually high Pauli limiting fields\cite{Bauer1}, helical vortex states\cite{Kaur} and even topologically protected states\cite{Sato}. Owing to these, since the discovery of the first non-centrosymmetric superconductor CePt$_3$Si\cite{Bauer1,Samokhin}, the study of such superconductors gained significant attention of the condensed matter physics community\cite{Sigrist,Hillier,Bauer2}.\\

Despite a number of theoretical predictions of the possibility of the exotic superconducting phases in non-centrosymmetric superconductors as discussed above, there has been no clear spectroscopic evidence of unconventionality in such superconductors studied till date. In this paper, we report our ultra-low temperature scanning tunneling microscopy and spectroscopy results on a non-centrosymmetric superconductor Ru$_7$B$_3$. Ru$_7$B$_3$ belongs to the space group $P6_3mc$ and the cyclic crystallographic class $C_{6v}$\cite{Aronsson}. Matthias $et$ $al.$ had first reported superconductivity in Ru$_7$B$_3$ in 1961\cite{Matthias}. However, owing to its low critical temperature, the system did not find much interest among the superconductivity community. Almost three decades later, the absence of the inversion symmetry in its crystal structure was highlighted by Morniroli $et$ $al.$\cite{Morniroli}. In various transport and thermodynamic measurements in the past\cite{Kase, Fang,Singh}, it was seen that $\Delta C_e/(\gamma_nT_c)$  and 2$\Delta_0/(k_B T_c$) in Ru$_7$B$_3$ were approximately 1.4 and 3.3 respectively indicating a weak-coupling superconducting state. These measurements also indirectly indicated that a predominant fully gapped $s$-wave order parameter could describe the superconducting state of Ru$_7$B$_3$ well. However, as we note, certain special features of the data presented in \cite{Kase, Fang} were ignored while making a claim for absence of unconventional superconductivity in Ru$_7$B$_3$. The most intriguing among such special features was a kink in the field dependent $\rho$-$T$ data\cite{Fang}, beyond which the superconducting transition curves split into two parts. The two parts exhibited significantly different sensitivity to the applied magnetic field and led to two dramatically different field scales for the upper critical fields ($H_{c2} \sim$ 1.1 T and 5 T respectively). In this context, Fang $et$ $al.$ discussed the possibility of a mixed angular momentum symmetry of the superconducting order parameter in Ru$_7$B$_3$. In addition, though it was ignored by the authors, a possible signature of unconventional pairing was also present in the specific heat data as presented in \cite{Kase}. More recently, Cameron $et$ $al.$ performed small-angle neutron scattering \cite{Cameron} on Ru$_7$B$_3$ and reported that the orientation of the vortex lattice in Ru$_7$B$_3$ strongly depends on the history of the applied magnetic field thereby indicating the possibility of a broken time-reversal symmetry in the order parameter. In order to probe the true order parameter symmetry of Ru$_7$B$_3$, we carried out detailed and direct temperature and magnetic field dependent scanning tunneling spectroscopy measurements on Ru$_7$B$_3$. The analysis of such data reveals spectroscopic signature of an order parameter with mixed angular momentum symmetry.\\

The single crystals used for our measurements showed a bulk superconducting transition at 2.6 K (Figure 1(b)). The scanning tunnelling microscopy (STM) and spectroscopy (STS) experiments were performed in a Unisoku system with RHK R9 controller, inside an ultra-high vacuum (UHV) cryostat at $\sim$ $10^{-10}$ mbar pressure. The lowest temperature down to which the measurements were performed was 300 mK. The STM is also equipped with a superconducting solenoid capable of producing a magnetic field up to 11 T. Since the single crystals could not be cleaved using the standard cleaving technique (optimized for layered materials only), we cleaned the surface by reversed sputtering for 30 minutes with Argon (Ar) ion \textit{in-situ} inside an integrated UHV preparation chamber. Following that, we immediately transferred the sample to the scanning stage at low temperature. The Tungsten (W) tip which was prepared outside by electrochemical etching was also cleaned in UHV by bombarding it with a high-energy electron-beam. This process helped us probe the pristine surface of Ru$_7$B$_3$. In the inset of Figure 1(b) we show an STM topographic image showing the distinctly visible crystallites with average grain size $\sim$ 3 nm. \\

In Figure 1(a) we show six representative tunneling spectra ($S_1$ - $S_6$) captured at randomly chosen points on the surface of Ru$_7$B$_3$ at $\sim$ 310 mK. A visual inspection reveals that based on their overall shapes, the spectra can be distinctively divided into two categories. The first type ($S_1$, $S_2$ and $S_3$) shows coherence peaks around $\pm$ 0.30 mV and shallow ‘V’-shaped dip in between the peaks. The spectra ($S_4$, $S_5$ and $S_6$) of second type show coherence peaks at $\pm$ 0.47 mV and they have a higher curvature and depth ($dI/dV$ $\sim$ 0 at $V$=0)  below that. We have also analysed these spectra within a single band ‘$s$-wave’ model\cite{Bardeen} using Dyne's formula\cite{Dynes}:
$N_s(E) \propto Re\left(\frac{(E-i\Gamma)}{\sqrt{(E-i\Gamma)^2-\Delta^2}}\right)$. The tunnelling current is given by $I(V) \propto \int_{-\infty}^{+\infty} N_s(E)N_n(E-eV)[f(E)-f(E-eV)]dE$. Here, $N_s(E)$ and $N_n(E)$ are the density of states (DOS) of the superconducting sample and the normal metallic tip respectively, while $f(E)$ is the Fermi-Dirac distribution function. $\Gamma$ is the Dyne's parameter that takes care of broadening of DOS. The theoretical plots thus generated are shown as black lines on the experimental data points in the Figure. It is seen that the spectra $S_1$, $S_2$ and $S_3$ give $\Delta$ of the order of 0.31 meV, 0.29 meV and 0.24 meV respectively while spectra $S_4$, $S_5$ and $S_6$ provide the values 0.48 meV, 0.42 meV and 0.49 meV respectively for the same analysis. It is also noted that, while the first group of spectra ($S_1$, $S_2$ and $S_3$) shows reasonably good fitting with the single gap ‘$s$-wave’ model (albeit with large $\Gamma$), the second group ($S_4$, $S_5$ and $S_6$) exhibits a significant departure from that.\\

The above-mentioned discrepancy between the experimental spectra and the spectra generated theoretically within a single-band ‘$s$-wave’ model, prompted us to consider other possible symmetries of the order parameter. To perform such an analysis, we modified Dyne's equation by introducing a more general expression\cite{Tanaka} of $\Delta(\theta)$ than an isotropic $\Delta$. The modified Dyne's equation reads as $N_s(E,\theta) \propto Re\left(\frac{(E-i\Gamma)}{\sqrt{(E-i\Gamma)^2-(\Delta' Cos(n\theta))^2}}\right)$. Here, $\theta$ is the polar angle (w.r.t. $(001)$) and the integer $n$ can be 0, 1 or 2 for $s$, $p$ and $d$ wave symmetries respectively. The expression for tunnelling current is also modified to $I(V) \propto \int_{-\infty}^{+\infty}\int_{0}^{2\pi} N_s(E,\theta)N_n(E-eV)[f(E)-f(E-eV)]d\theta dE$. In Figure 1(c) we show the experimental spectrum $S_4$ along with theoretical plots considering isotropic ‘$s$-wave’ $\Delta$ (red line) and anisotropic ‘$p$-wave’ $\Delta$ (blue line). It is clear that the spectrum, especially the ‘V’-shaped part of that between the coherence peaks, is better described by the ‘$p$-wave’ symmetry. It is also interesting to note that, the extracted value of $\Delta$ (0.47 meV) for such fit does not differ much with the same from the best ‘$s$-wave’ fit (0.48 meV). Such ‘V’-shaped spectra are often seen for superconductors with possible unconventional symmetries and are well described by the Tanaka-Kashiwaya model\cite{Tanaka} we used here. For example, from tunnelling spectroscopic study on SmFeAsO$_{0.85}$, Millo $et$ $al.$\cite{Millo} associated such a shape with an unconventional order parameter. To note, it was also reported there that some of the STM spectra could also be fitted well within pure ‘$s$-wave’ model but with significantly smaller $\Delta$ and relatively higher $\Gamma$ -- a situation similar to our group-I spectra.\\

Now we focus on the spectrum $S_1$, which belongs to group I and fits reasonably well with single ‘$s$-wave’ gap (and with relatively large $\Gamma$). A closer inspection, however, reveals that there is a small discrepancy between the experimental data and the ‘$s$-wave’ model spectrum. We investigated the evolution of this discrepancy with temperature. Temperature dependence of $\Delta$ approximately followed the BCS prediction (Figure 2(c)) with $\Delta_0$ = 0.3 meV. The broadening parameter $\Gamma$ did not change much within this range. The departure of the experimental spectrum from the $s$-wave model rapidly decreased with increasing temperature and at around 750 mK the discrepancy almost disappeared (Figure 2(a)). To illustrate this effect clearly, we show two spectra, one at 338 mK and another at 961 mK along with their theoretical ($s$-wave) fits in Figure 2(b). This observation indicates the possibility of a mixed angular momentum symmetry in the order parameter where the amplitudes corresponding to different $l$ vary differently with temperature. This also explains why Fang $et$ $al.$ did not find any signature of unconventionality in their lower critical field ($H_{c1}$) studies\cite{Fang} down to 1.2 K, which is well above the temperature window where we could notice deviation from $s$-wave behaviour in our data. We have also performed magnetic field dependence of another spectrum ($S_3$) from the same group I (Figure 3(a)). The key observation is that while the spectra deviate appreciably from ‘$s$-wave’ theoretical curves at low fields, above 10 kG the spectra resembled far more closely to the ‘$s$-wave’ predictions. For demonstrating this effect more clearly, we show the spectra and respective ‘$s$-wave’ fits at zero field and at 16 kG in Figure 3(b). The evolution of the extracted gap ($\Delta$) and broadening parameter ($\Gamma$) with the applied magnetic field are shown in Figure 3(c).\\

In Figure 4(a) we demonstrated the detailed temperature dependent study of spectra $S_6$ which belong to group II. At $\sim$ 740 mK a small zero-bias conductance peak (ZBCP) appeared. It became more and more pronounced up to $\sim$ 1370 mK and then started fading until it, along with other prominent spectral features, gradually disappeared at 1.9 K. To note, emergence of such a ZBCP can be attributed to a ‘$p$-wave’ component in the order parameter where the interface normal and the lobe-direction of the ‘$p$-wave’ maintain an acute angle between them\cite{Yamashiro, Laube}. Since the surface of our Ru$_7$B$_3$ has crystallites with random orientations and the tip is engaged randomly at different points, this condition can naturally be satisfied some times.\\

Based on the discussion above, if Ru$_7$B$_3$ has a mixed angular momentum symmetry in its order parameter, then we would expect the unconventional component to be present in group-I spectra too. To understand this aspect in details, we used an ‘$s+p$-wave’ model to analyse the spectra $S_1$. In this model, the effective gap is given by $\Delta_{s+p}$ = $\Delta_s$ + $\Delta_p Cos\theta$. In Figure 4(b) we present the spectrum $S_1$ (black circles) at 338 mK with numerically generated spectra considering ‘$s+p$-wave’ symmetry (blue line). Pure ‘$s$-wave’ fit (red line) is also shown for comparison. It is clear that the mixed angular momentum symmetry provides a better description of the data. Temperature evolutions for the amplitudes of the two components $\Delta_s$ and $\Delta_p$ extracted from the above ‘$s+p$-wave’ fittings for the whole spectra $S_1$ are presented in Figure 4(c). Not as a surprise, while the convensional $\Delta_s$ follows a smooth BCS like dependence upto 2 K, the smaller $\Delta_p$ sharply drops and goes beyond our measurement resolution at 0.9 K.\\ 

It should be noted here that the superconducting gap is the manifestation of a phase-coherent macroscopic condensate. Therefore, the measured $\Delta$ should ideally be unique irrespective of the measurement techniques. Since the system is metallic in nature, any role of special surface states is ruled out. {The $T_c$ and the $H_c$ that we measure for all our spectra match well irrespective of whether they fall under Group I or II.} Therefore, it can be concluded that, all the spectra falling under two distinct groups differ with each other based on which component of the order parameter contributes predominantly for a particular crystallite that the measurement is performed on.\\

In summary, we performed scanning tunnelling spectroscopy on single-crystal Ru$_7$B$_3$ and recorded several spectra which can be broadly categorised into two groups, group-I and II. Group-I consists of spectra which are shallow in shape. They show overall good agreement with single gap ‘$s$-wave’ symmetry except at very low temperatures where they deviate from such pure $s$-wave model. Smaller superconducting gap and larger broadening parameter are characteristics of these spectra. Group-II spectra are broader in shape and show sharper coherence peaks. They significantly deviate from the predictions of ‘$s$-wave’ model but a theoretical model with ‘$p$-wave’ symmetry shows better agreement. The superconducting gap is larger, and the broadening parameter is very small for such spectra compared to those belonging to the first group. The temperature dependences of the spectra belonging to both the groups indicate the presence of a mixed ‘$s+p$-wave’ symmetry in the order parameter where the two components have different temperature dependence.\\

We thank Tanmoy Das for his useful comments. R.P.S. acknowledges the financial support from the Science and Engineering Research Board (SERB)-Core Research Grant (grant No. CRG/2019/001028). G.S. would like to acknowledge financial support from the Swarnajayanti fellowship awarded by the Department of Science and Technology (DST), Govt. of India (grant No. DST/SJF/PSA-01/2015-16).

\begin{figure}[h!]
	\centering
	\includegraphics[width=1\textwidth]{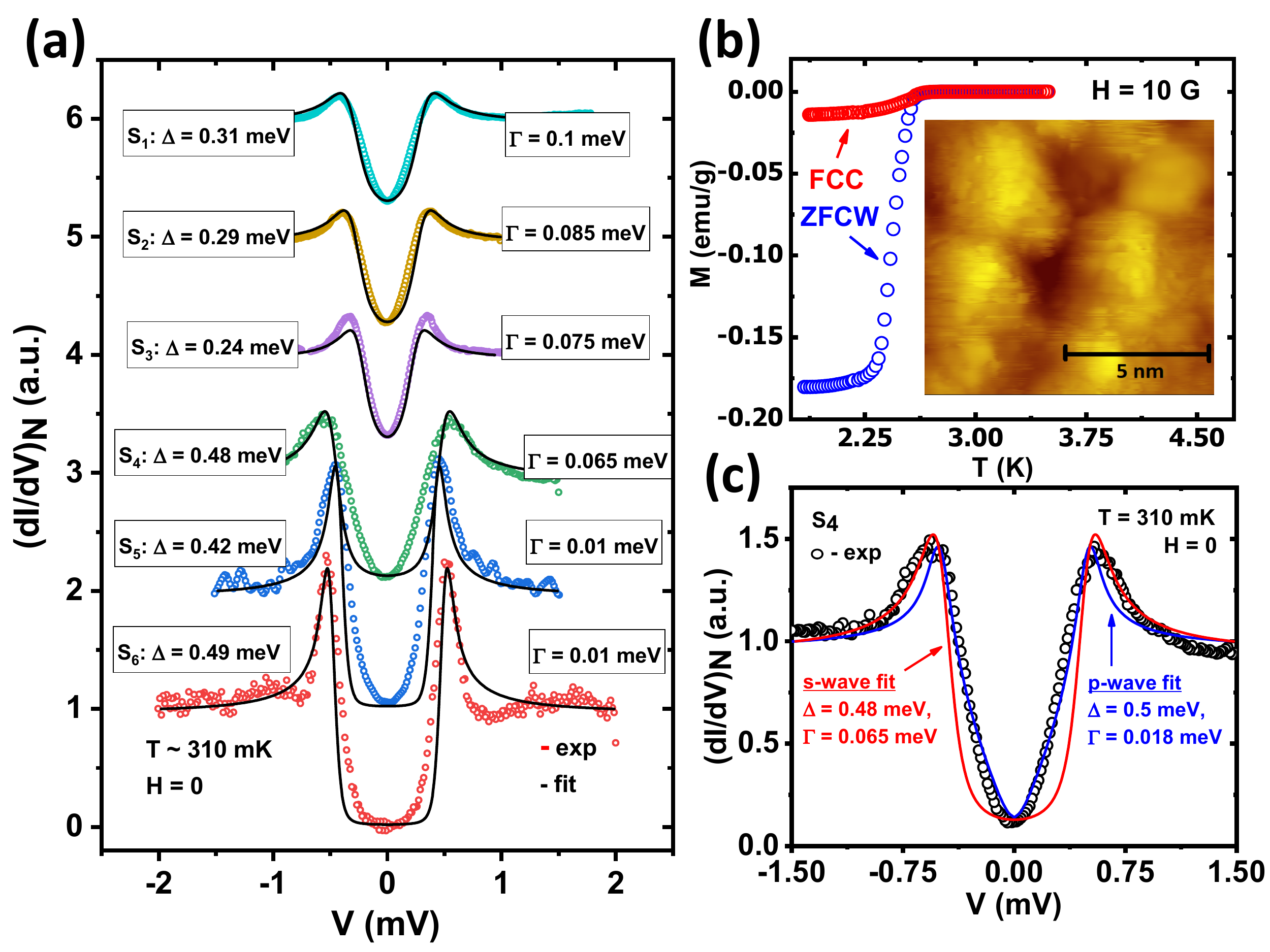}
	\caption{\textbf{(a)} Six representative tunnelling spectra ($S_1$ - $S_6$) plots (colour points) along with corresponding numerically generated spectra under single gap $s$-wave model (black lines). The extracted fitting parameters $\Delta$ and $\Gamma$ are shown also for each spectrum. \textbf{(c)} Spectrum $S_4$ with both best single gap ‘$s$-wave’ and ‘$p$-wave’ fit alongwith corresponding extracted parameters. The temperature ($T$) $\sim$ 310 mK for all spectra. \textbf{(b)} Bulk magnetization ($M$) data in both zero field cool warming (ZFCW) and field cool cooling (FCC) condition with 10 G magnetic field. $inset$: STM topographic image of the sample.}
	
\end{figure}

\begin{figure}[h!]
	\centering
	\includegraphics[width=1\textwidth]{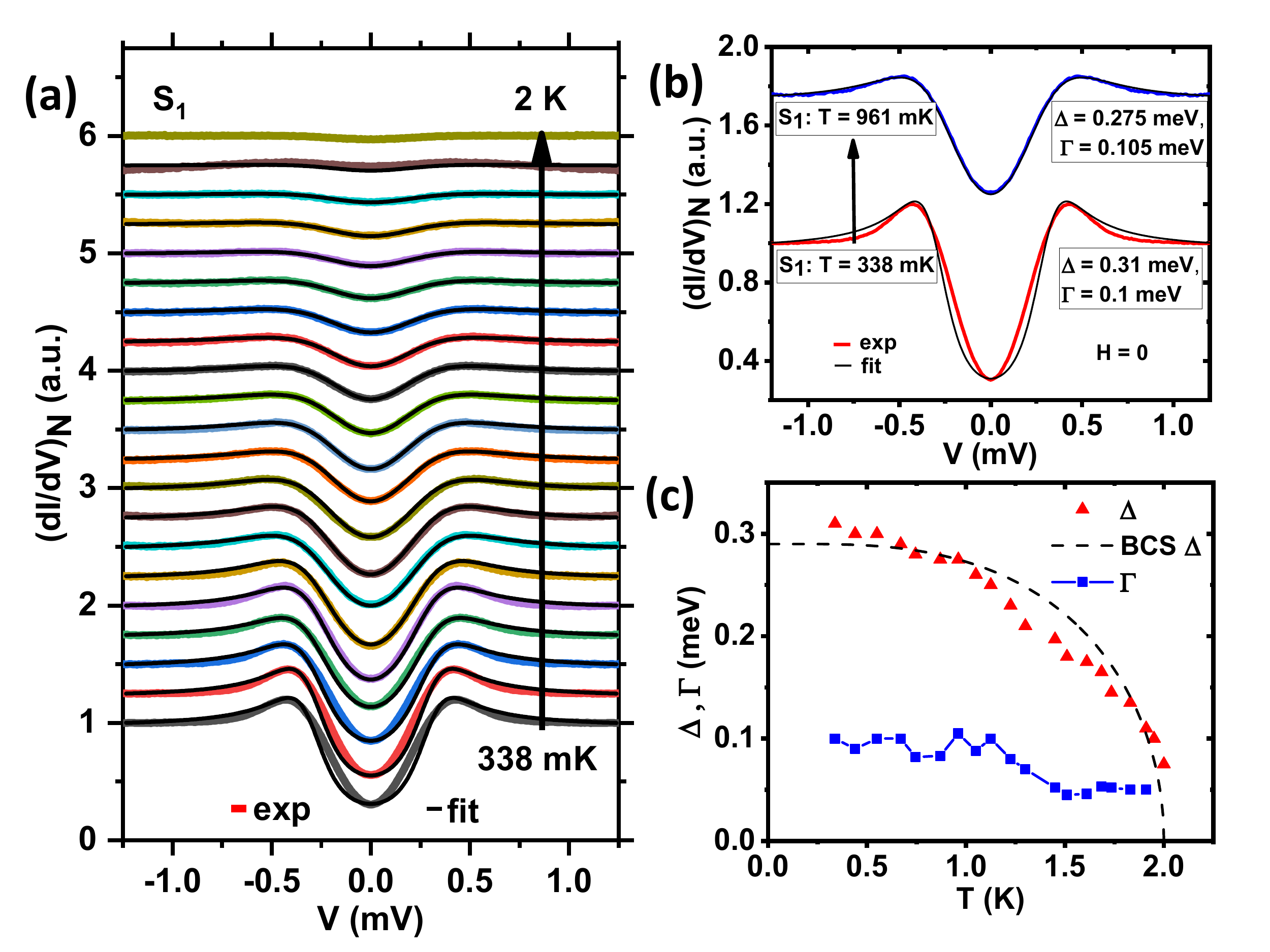}
	\caption{\textbf{(a)} Temperature (T) dependence of tunnelling conductance spectra $S_1$ (colour lines) with theoretical fits (black lines) in absence of any magnetic field. \textbf{(b)} Spectra $S_1$ at 340 mK and also at 960 mK along with corresponding fitting parameters, where better fit at higher temeperature is visible. \textbf{(c)} Evolution of $\Delta$ and $\Gamma$ with temperature, extracted from the plot (a) along with an ideal BCS trend of $\Delta$ for comparison.}
	
\end{figure}

\begin{figure}[h!]
	\centering
	\includegraphics[width=1\textwidth]{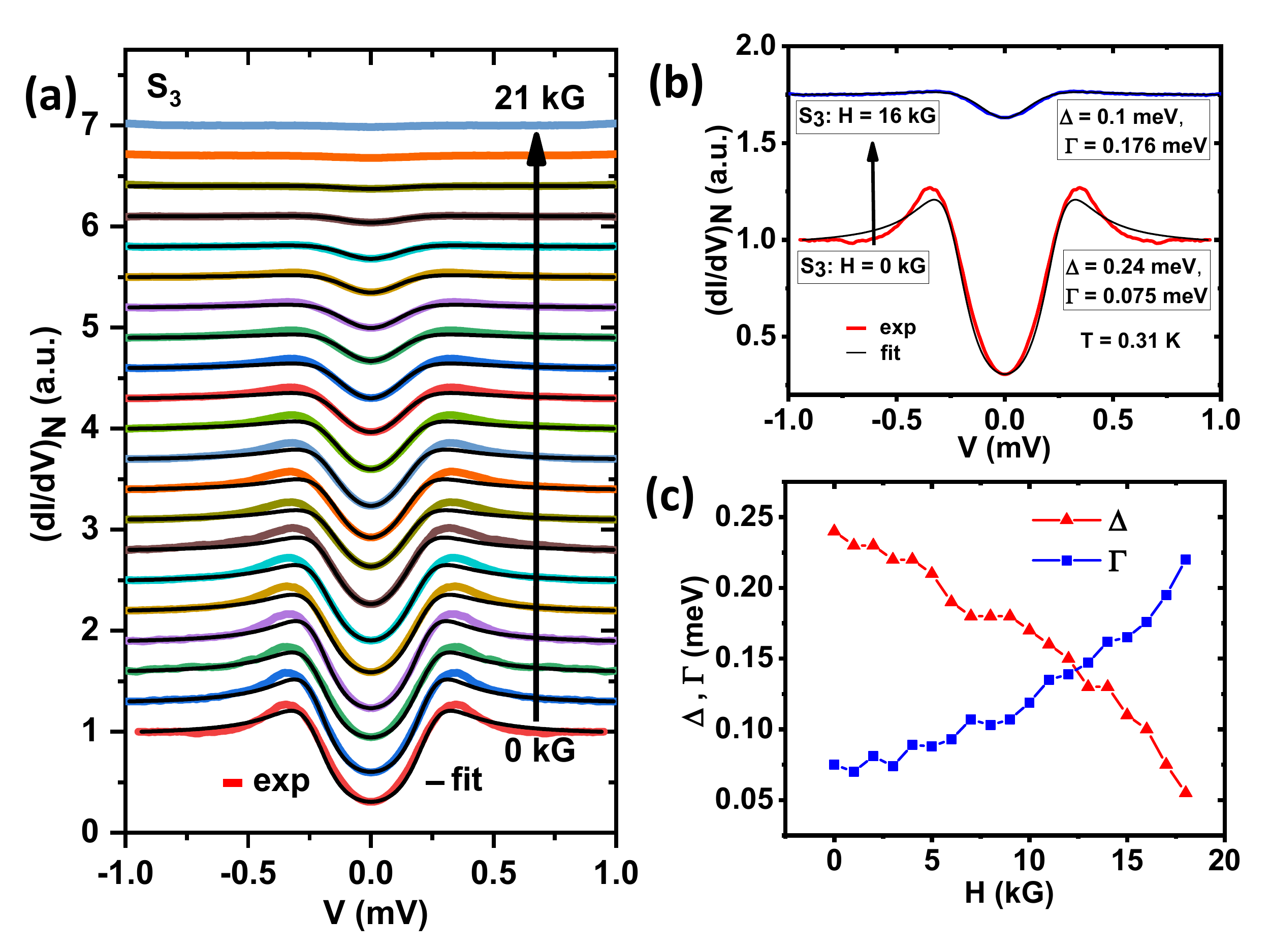}
	\caption{\textbf{(a)} Magnetic field ($H$) dependence of tunnelling conductance spectra $S_3$ (colour lines) with theoretical fits (black lines) all measured at T $\sim$ 310 mK. \textbf{(b)} Spectra $S_3$ in the environment of $H$ = 0 and $H$ = 16 kG field, along with corresponding fitting parameters. $H\parallel c$-axis of the crystal and a better fit at higher field is visible. \textbf{(c)} Evolution of $\Delta$ and $\Gamma$ with magnetic field, extracted from the plot (a).}
	
\end{figure}

\begin{figure}[h!]
	\centering
	\includegraphics[width=1\textwidth]{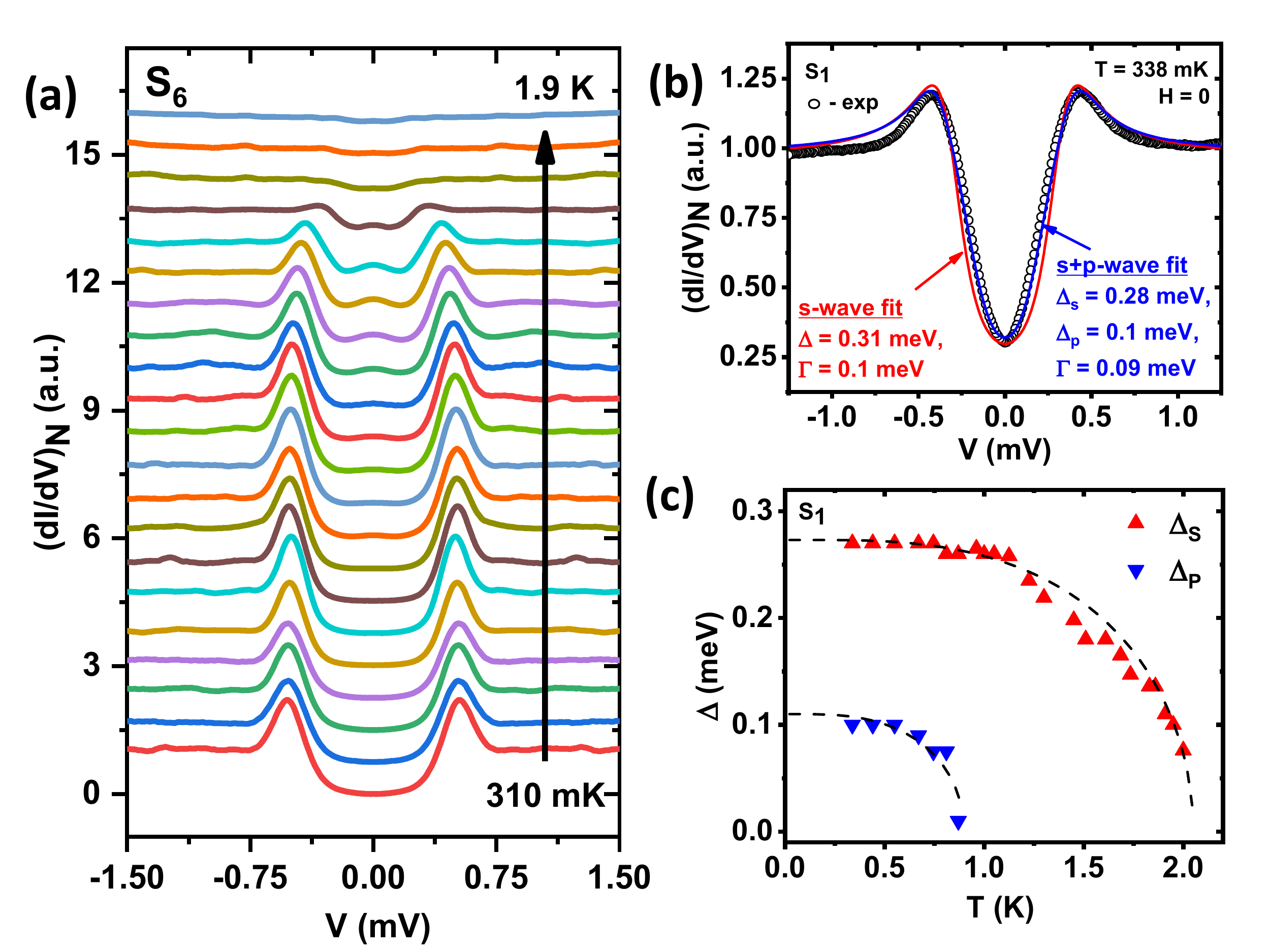}
	\caption{\textbf{(a)} Temperature (T) dependence of tunnelling conductance spectra $S_6$ incorporating gradual appearance and disappearance of the peak like feature. \textbf{(b)} Spectrum $S_1$ along with best pure $s$-wave (red line) and mixed $s+p$-wave (blue line) fits with corresponding extracted parameters $\Delta$ and $\Gamma$. \textbf{(c)} Evolution of $\Delta_s$ and $\Delta_p$ with temperature, extracted from the $s+p$-wave fits of spectra $S_1$. Ideal BCS trends of $\Delta$ are also shown for comparison.}
	
\end{figure}

\end{document}